\documentclass[pra,aps,showpacs,twocolumn]{revtex4}
\usepackage{amssymb}

\usepackage{graphicx}
\usepackage{latexsym}
\usepackage{amsmath}
\usepackage[dvipdfm]{hyperref}
\usepackage{float}

\begin{document}

\title{Lazy states, discordant states and entangled states for 2-qubit systems}
\author{Jianwei Xu}
\email{xxujianwei@126.com}
\affiliation{College of Science, Northwest A\&F University, Yangling, Shaanxi 712100, China}
\date{\today}

\begin{abstract}
We investigate the lazy states, entangled states and discordant states for
2-qubit systems. We show that many lazy states are discordant, many lazy
states are entangled, and many mixed entangled states are not lazy. With
these investigations, we provide a laziness-discord-entanglement hierarchy
diagram about 2-qubit quantum correlations.
\end{abstract}

\pacs{03.65.Ud, 03.67.Mn, 03.65.Aa}
\maketitle

\section{Introduction}

Quantum correlation is one of the most striking features of quantum theory.
Entanglement is the most famous kind of quantum correlation, and leads to
powerful applications \cite{Horodecki2009}. Discord is another kind of quantum
correlation, which captures more correlation than entanglement in the sense
that a disentangled state may have no zero discord \cite{Modi2012}. Due to the
theoretical and applicational interests, discord has been extensively studied
\cite{Modi2012} and still in active research (for examples see \cite{Rulli2011,Xu2013,Chi2013,Liu2013}).

A bipartite state is called lazy, if the entropy rate of one subsystem is
zero under any coupling to the other subsystem. Necessary and sufficient
conditions have recently been established for a state to be lazy
\cite{Rosario2011}, and it was shown that almost all states are pretty
lazy \cite{Hutter2012}. It is shown that a maximally entangled pure state is lazy\cite{Ferraro2010}. This indicates that the correlation described by lazy states is not the same by entanglement. So we are
interested to clarify the question that, whether there are many lazy states
which are entangled, and whether there are many entangled states which are
lazy. This paper answers this question for the 2-qubit case.

This paper is organized as follows. In Section 2, we briefly review the
definitions of entangled states, discordant states and lazy states. In
Section 3, we establish a necessary and sufficient condition for 2-qubit
lazy states. In Section 4, we show that there are many 2-qubit lazy states
which are discordant states. In Section 5, we show that there are many
disentangled states which are not lazy. In Section 6, we show that there are
many 2-qubit mixed lazy states which are entangled. In section 7, we briefly
summary this paper by providing a laziness-discord-entanglement hierarchy
diagram to characterize the bipartite quantum correlations.

\section{Entangled states, discordant states, lazy states}

We briefly review the definitions about entangled states, discordant states
and lazy states.

Finite-dimensional quantum systems $A$ and $B$ are described by the Hilbert
spaces $H^{A}$ and $H^{B}$ respectively, the composite system $AB$ is then
described by the Hilbert space $H^{A}\otimes H^{B}$. Let $n_{A}=\dim H^{A}$%
, $n_{B}=\dim H^{B}$. A state $\rho ^{AB}$ is called a disentangled state
(or separable state) if it can be written in the form\bigskip
\begin{gather}
\rho ^{AB}=\sum_{i}p_{i}\rho _{i}^{A}\otimes \rho _{i}^{B},
\end{gather}
where $p_{i}\geq 0,\sum_{i}p_{i}=1,\{\rho _{i}^{A}\}_{i}$ are density
operators on $H^{A}$, $\{\rho _{i}^{B}\}_{i}$ are density operators on $%
H^{B}.  $If $\rho ^{AB}$ is disentangled we then say $E(\rho ^{AB})=0.$

A state $\rho ^{AB}$ is called a zero-discord state with respect to $A$ if
it can be written in the form\bigskip
\begin{gather}
\rho ^{AB}=\sum_{i=1}^{n_{A}}p_{i}|\psi _{i}^{A}\rangle \langle \psi
_{i}^{A}|\otimes \rho _{i}^{B},
\end{gather}
where $p_{i}\geq 0,\sum_{i}p_{i}=1,\{|\psi _{i}^{A}\rangle \}_{i}$ is an
orthonormal basis for $H^{A}$, $\{\rho _{i}^{B}\}_{i}$ are density operators
on $H^{B}.  $If $\rho ^{AB}$ is in the form Eq.(2) we then say $D_{A}(\rho
^{AB})=0.$

Evidently,
\begin{gather}
D_{A}(\rho ^{AB})=0 \ ^{\Rightarrow }  _{\nLeftarrow } \ E(\rho ^{AB})=0.
\end{gather}

A state $\rho ^{AB}$ is called a lazy state with respect to $A$ if \cite{Rosario2011}
\begin{gather}
C_{A}(\rho ^{AB})=[\rho ^{AB},\rho ^{A}\otimes I^{B}]=0,
\end{gather}
where $\rho ^{A}=tr_{B}\rho ^{AB}$, $I^{B}$ is the identity operator on $%
H^{B}.$ An important physical interpretation of lazy states is that the
entropy rate of $A$ is zero in the time evolution under any coupling to $B,$
\begin{gather}
C_{A}(\rho ^{AB}(t))=0\Leftrightarrow \frac{d}{dt}tr_{A}[\rho ^{A}(t)\log
_{2}\rho ^{A}(t)]=0\text{.}
\end{gather}

$D_{A}(\rho ^{AB})=0$ and $C_{A}(\rho ^{AB})=0$ has the inclusion relation
below \cite{Ferraro2010}
\begin{gather}
D_{A}(\rho ^{AB})=0 \ ^{\Rightarrow} _{\nLeftarrow} \ C_{A}(\rho ^{AB})=0.
\end{gather}
Maximal pure entangled states are the examples of $C_{A}(\rho ^{AB})=0$ but $%
D_{A}(\rho ^{AB})\neq 0$ \cite{Ferraro2010}.

The direct product states have the form
\begin{gather}
\rho ^{AB}=\rho ^{A}\otimes \rho ^{B},
\end{gather}
they are obviously zero-discord states.

\section{The form of 2-qubit lazy states}

Any 2-qubit state can be written in the form \cite{Fano1983}
\begin{gather}
\rho ^{AB}=\frac{1}{4}(I\otimes I+\sum_{i=1}^{3}x_{i}\sigma _{i}\otimes
I+\sum_{j=1}^{3}y_{j}I\otimes \sigma _{j}  \notag \\
+\sum_{i,j=1}^{3}T_{ij}\sigma
_{i}\otimes \sigma _{j}),
\end{gather}
where $I$ is the two-dimensional identity operator,$\{\sigma
_{i}\}_{i=1}^{3} $ are Pauli operators, $\{x_{i}\}_{i=1}^{3},\{y_{j}%
\}_{j=1}^{3},\{T_{ij}\}_{i,j=1}^{3},$ are all real numbers satisfying some
conditions (we will explore these conditions when we need them) to ensure
the positivity of $\rho ^{AB}$, $\rho ^{A}$ and $\rho ^{B}$. We often omit $%
I $ for simplicity without any confusion.

$ $

\textbf{Proposition 1.}
The 2-qubit state $\rho ^{AB}$ in Eq.(8) is lazy if and only if
\begin{gather}
\{x_{i}\}_{i=1}^{3} // \{T_{ij}\}_{i=1}^{3} \text{ for }j=1,2,3.
\end{gather}

$ $

\textbf{Proof.} For state in Eq.(8),
\begin{gather}
\rho ^{A}=\frac{1}{2}(I+\sum_{k=1}^{3}x_{k}\sigma _{k}\otimes I), \\
[\rho ^{AB},\rho ^{A}]=\frac{1}{8}\sum_{ijk=1}^{3}T_{ij}x_{k}[\sigma _{i}\otimes
\sigma _{j},\sigma _{k}\otimes I] \notag \\
=\frac{1}{8}\sum_{ijk=1}^{3}T_{ij}x_{k}[\sigma _{i},\sigma _{k}]\otimes
\sigma _{j} \notag \\
=\frac{i}{4}\sum_{ijkl=1}^{3}T_{ij}x_{k}\varepsilon _{ikl}\sigma
_{l}\otimes \sigma _{j}.
\end{gather}
In the last line, $\varepsilon _{ikl}$ is the permutation symbol.

Let $[\rho ^{AB},\rho ^{A}]=0,$ then
\begin{gather}
\sum_{ik=1}^{3}T_{ij}x_{k}\varepsilon _{ikl}=0,
\end{gather}
this evidently leads to Eq.(9). $\square $

\section{Lazy but diacordant 2-qubit states}

It is easy to check that $C_{A}(\rho ^{AB})=0$ defined in Eq.(4) is invariant
under locally unitary transformations for arbitrary $n_{A}$ and $n_{B}$.
Under locally unitary transformations, any 2-qubit state in Eq.(8) can be
written in the form \cite{Luo2008}
\begin{gather}
\rho ^{AB}=\frac{1}{4}(I\otimes I+\sum_{i=1}^{3}x_{i}\sigma _{i}\otimes
I+\sum_{j=1}^{3}y_{j}I\otimes \sigma _{j}     \notag \\
+\sum_{i=1}^{3}\lambda _{i}\sigma
_{i}\otimes \sigma _{i}),
\end{gather}
where $0\leq \lambda _{1}\leq \lambda _{2}\leq \lambda _{3}$ being the
singular values of $\{T_{ij}\}_{ij}$ in Eq.(8). Note that $%
\{x_{i}\}_{i=1}^{3},\{y_{j}\}_{j=1}^{3}$ in Eq.(9) are not the same with in
Eq.(8).

We now look for the conditions such that $D_{A}(\rho ^{AB})=0.$  Suppose
$D_{A}(\rho ^{AB})=0$, then according to Eq.(2), there exists real vector $%
\overrightarrow{n}=\{n_{1},n_{2},n_{3}\}$ with $%
n_{1}^{2}+n_{2}^{2}+n_{3}^{2}=1$ such that
\begin{gather}
\rho ^{AB}=\Pi _{0}\otimes I\rho ^{AB}\Pi _{0}\otimes I+\Pi _{1}\otimes
I\rho ^{AB}\Pi _{1}\otimes I,
\end{gather}
with
\begin{gather}
\Pi _{0}=\frac{1}{2}(I+\overrightarrow{n}\cdot \overrightarrow{\sigma }), \\
\Pi _{1}=\frac{1}{2}(I-\overrightarrow{n}\cdot \overrightarrow{\sigma }).
\end{gather}
It can be check that
\begin{gather}
\Pi _{0}\sigma _{i}\Pi _{0}+\Pi _{1}\sigma _{i}\Pi _{1}=n_{i}
\overrightarrow{n}\cdot \overrightarrow{\sigma }.
\end{gather}
Then Eq.(14) becomes
\begin{gather}
\rho ^{AB}=\frac{1}{4}(I\otimes I+\sum_{i=1}^{3}x_{i}n_{i}\overrightarrow{n}%
\cdot \overrightarrow{\sigma }\otimes I   \notag \\
+\sum_{j=1}^{3}y_{j}I\otimes \sigma
_{j}+\sum_{i=1}^{3}\lambda _{i}n_{i}\overrightarrow{n}\cdot \overrightarrow{%
\sigma }\otimes \sigma _{i})   \notag \\
=\frac{1}{4}(I\otimes I+\sum_{ij=1}^{3}x_{i}n_{i}n_{j}\sigma _{j}\otimes I    \notag \\
+\sum_{j=1}^{3}y_{j}I\otimes \sigma _{j}+\sum_{ij=1}^{3}\lambda
_{i}n_{i}n_{j}\sigma _{j}\otimes \sigma _{i}).
\end{gather}
Comparing to Eq.(13), then for $j=1,2,3,$
\begin{gather}
\sum_{i=1}^{3}x_{i}n_{i}n_{j}=x_{j}\Rightarrow \overrightarrow{n}//%
\overrightarrow{x},  \\
\lambda _{i}n_{i}n_{j}=\delta _{ij}\lambda _{j}=\delta _{ij}\lambda _{i}
\Rightarrow \lambda _{i}=0 \ \text{or} \  n_{i}=\pm 1.
\end{gather}
(i).If $\lambda _{1}=\lambda _{2}=\lambda _{3}=0,$ let $\overrightarrow{n}//%
\overrightarrow{x},$ then $D_{A}(\rho ^{AB})=0$.

(ii).If $0=\lambda _{1}=\lambda _{2}<\lambda _{3}=0,$ then $\overrightarrow{n}%
=(0,0,\pm 1),$ to satisfy $\overrightarrow{n}//\overrightarrow{x},$ we see that only when $%
\overrightarrow{x}=(0,0,x_{3})$ we have
$D_{A}(\rho ^{AB})=0$.

(iii).If $0=\lambda _{1}<\lambda _{2}<\lambda _{3}=0,$ then Eq.(20) can not be
satisfied, so $\rho ^{AB}$ is discordant.

(iv).If $0<\lambda _{1}<\lambda _{2}<\lambda _{3}=0,$ then Eq.(20) can not be
satisfied, so $\rho ^{AB}$ is discordant.

Comparing with Proposition 1, we then get Proposition 2 below.

$ $

\textbf{Proposition 2.} A 2-qubit state in Eq.(13) is lazy but discordant if and only
if $\overrightarrow{x}=0$ and $0<\lambda _{2}<\lambda _{3}$.

$ $

Since any locally unitary transformation keeps $\overrightarrow{x}=0$ invariant in Eq.(8), then we rewrite Proposition 2 as Proposition 2$'$ below.

$ $

\textbf{Proposition 2$'$.} A 2-qubit state in Eq.(8) is lazy but discordant if and only
if $\overrightarrow{x}=0$ and the matrix $\{T_{ij}\}_{ij}$ have at least two positive singular values.

$ $

We make a note that some constraints about $\{y_{j}\}_{j=1}^{3},\lambda
_{1},\lambda _{2},\lambda _{3}$ are required to guarantee the positivity of $%
\rho ^{AB},\rho ^{A}$, $\rho ^{B}$ in Proposition 2$.$These constraints are
rather complex since there are so many parameters. To show there indeed
exist many states described in Proposition 2, we choose some special states.
For the state
\begin{gather}
\rho ^{AB}=\frac{1}{4}(I\otimes I+\sum_{j=1}^{3}y_{j}I\otimes \sigma
_{j}+\sum_{i=1}^{3}\lambda _{i}\sigma _{i}\otimes \sigma _{i}),
\end{gather}
where $0\leq \lambda _{1}\leq \lambda _{2},0<\lambda _{2}<\lambda _{3},$ we have $\rho ^{A}=I$,and
\begin{gather}
\rho ^{B}=\frac{1}{2}(I+\sum_{j=1}^{3}y_{j}\sigma _{j}).
\end{gather}
$\rho ^{B}$ is positive then
\begin{gather}
\sum_{j=1}^{3}y_{j}^{2}\leq 1.
\end{gather}
Let $y_{2}=y_{3}=\lambda _{1}=0,$ then the four eigenvalues of $\rho ^{AB}$
in Eq.(21) are
\begin{gather}
\frac{1}{4}(1\pm \sqrt{y_{1}^{2}+(\lambda _{3}\pm \lambda _{2})^{2}}).
\end{gather}
These eigenvalues are all nonnegtive then we need
\begin{gather}
0<\lambda _{2}<\lambda _{3},   \\
y_{1}^{2}+(\lambda _{3}+\lambda _{2})^{2}\leq 1.
\end{gather}
There are many triples $\{y_{1},\lambda _{3},\lambda _{2}\}$ satisfy Eqs.(25,26),
then the corresponding states in Eq.(21) are lazy but discordant states.

\section{Some disentangled but not lazy 2-qubit states}

To show there exist many 2-qubit states which are disentangled but not lazy,
we consider the states of the form
\begin{gather}
\rho ^{AB}=p|\psi _{1}^{A}\rangle \langle \psi _{1}^{A}|\otimes \rho
_{1}^{B}+(1-p)|\psi _{2}^{A}\rangle \langle \psi _{2}^{A}|\otimes \rho
_{2}^{B},
\end{gather}
where $p\in (0,1),\{|\psi _{i}^{A}\rangle \}_{i=1}^{2}$ are normalized
states in $H^{A}$ but not necessarily orthogonal,$\{\rho
_{i}^{B}\}_{i=1}^{2} $ are density operators on $H^{B}.$ Note that $p=0$ or $%
p=1$ leads to direct product states, so we do not consider such cases.

Under locally unitary transformations, we let
\begin{gather}
|\psi _{1}^{A}\rangle \langle \psi _{1}^{A}|=\frac{I+(0,0,1)\cdot
\overrightarrow{\sigma }}{2},   \\
|\psi _{2}^{A}\rangle \langle \psi _{2}^{A}|=%
\frac{I+(\sin \alpha ,0,\cos \alpha )\cdot \overrightarrow{\sigma }}{2},   \\
\rho _{1}^{B}=\frac{I+a(0,0,1)\cdot \overrightarrow{\sigma }}{2},    \\
\rho
_{2}^{B}=\frac{I+b(\sin \beta ,0,\cos \beta )\cdot \overrightarrow{\sigma }}{%
2},
\end{gather}
where $\alpha ,\beta \in \lbrack 0,\pi ],a,b\in \lbrack 0,1].$

Some special states can be apparently specified.

(v).$\alpha =0,\rho ^{AB}$ in Eq.(27) are direct product states;

(vi).$\alpha =\pi ,\rho ^{AB}$ in Eq.(27) are zero-discord states;

(vii).$a=b=0,\rho ^{AB}$ in Eq.(27) are direct product states.

Now we consider the cases excluding (v), (vi), (vii) above. Taking Eqs.(28-31) into
Eq.(27), and using the notations in Eq.(8), we get
\begin{gather}
\overrightarrow{x}=((1-p)\sin \alpha ,0,p+(1-p)\cos \alpha ),  \\
\{T_{i1}\}_{i}=(b(1-p)\sin \alpha \sin \beta ,0,b(1-p)\cos \alpha \sin \beta ),  \\
\{T_{i2}\}_{i}=(0,0,0),  \\
\{T_{i3}\}_{i}=(b(1-p)\sin \alpha \cos \beta ,0,ap+b(1-p)\cos \alpha \cos \beta ).
\end{gather}
From Proposition 1, $\rho ^{AB}$ in Eq.(27) is lazy if and only if $%
\overrightarrow{x}//\{T_{i1}\}_{i}$ and $\overrightarrow{x}//\{T_{i3}\}_{i}.
$  Since $x_{1}=(1-p)\sin \alpha \neq 0,$  then  $\overrightarrow{x}//\{T_{i1}\}_{i}$ and $\overrightarrow{x}//\{T_{i3}\}_{i}$ lead to
\begin{gather}
b\sin \beta =0.  \\
a=b\cos \beta .
\end{gather}
Eqs.(36,37) together correspond to direct product states since $\rho _{1}^{B}=\rho
_{2}^{B} $. Otherwise, there are many states violate Eq.(36) or Eq.(37), so they
are not lazy states.

$ $

\textbf{Proposition 3}. 2-qubit disentangled state $\rho ^{AB}$ in Eq.(27),
is a direct product state when  $|\psi _{1}^{A}\rangle =|\psi _{2}^{A}\rangle $ or $\rho
_{1}^{B}=\rho _{2}^{B}$, is a zero-discord state when $\langle \psi
_{1}^{A}|\psi _{2}^{A}\rangle =0$.
Otherwise, $\rho ^{AB}$ is not lazy.

\bigskip

\section{Some lazy but entangled states}

We know that a bipartite pure state is lazy only if under locally unitary transformations it can be written in the form  \cite{Rosario2011} $|\psi^{AB}\rangle=\frac{1}{\sqrt{s}}\sum_{i=1}^{s}|\psi^{A}_{i}\rangle|\psi^{B}_{i}\rangle$, where $\{|\psi _{i}^{A}\rangle \}_{i}$ are orthonormal sets in $H^{A}$, $\{|\psi _{i}^{B}\rangle \}_{i}$  are orthonormal sets in $H^{B}$, $s\leq \min \{n_{A},n_{B}\}$. When $s= \min \{n_{A},n_{B}\}$ it is maximally
entangled state.  In this section we look for
more 2-qubit mixed states which are lazy but entangled.

From Proposition 1, we know the following 2-qubit Bell-diagonal states are
lazy
\begin{gather}
\rho ^{AB}=\frac{1}{4}(I\otimes I+\sum_{i=1}^{3}\lambda _{i}\sigma
_{i}\otimes \sigma _{i}),
\end{gather}
where $\{\lambda _{i}\}_{i=1}^{3}$ are real numbers satisfying some
constraints to ensure the positivity of $\rho ^{AB}.$

In this section, for convenience, we do not assume $\{\lambda
_{i}\}_{i=1}^{3}$ are all nonnegative. We represent the states in Eq.(38) in the $(\lambda _{1},\lambda _{2},\lambda _{3})$ space.

The eigenvalues of $\rho ^{AB}$ in Eq.(38) are
\begin{gather}
\frac{1}{4}\{1-\lambda _{1}+\lambda _{2}+\lambda _{3},1+\lambda
_{1}-\lambda _{2}+\lambda _{3},   \notag \\
1+\lambda _{1}+\lambda _{2}-\lambda
_{3},1-\lambda _{1}-\lambda _{2}-\lambda _{3}\}.
\end{gather}
Then the positivity of $\rho ^{AB}$ requires that $\{\lambda
_{i}\}_{i=1}^{3} $ are in the tetrahedron (with its boundary) with the
vertices $(-1,-1,-1),(-1,1,1),(1,-1,1),(1,1,-1)$ in the $(\lambda _{1},\lambda _{2},\lambda _{3})$ space \cite{Horodecki1996}. Disentangled states in Eq.(38) are in the octahedron
(with its boundary) with the vertices $(\pm 1,0,0),(0,\pm 1,0),(0,0,\pm 1)$
\cite{Horodecki1996}. From Proposition 2, we know the zero-discord states in
Eq.(38) are only three line segments $(\lambda _{1},0,0)$ with $\lambda
_{1}\in \lbrack -1,1], (0,\lambda _{2},0)$ with $\lambda _{2}\in \lbrack -1,1]$%
, $(0,0,\lambda _{3})$ with $\lambda _{3}\in \lbrack -1,1].$

Then the states in the tetrahedron (with its boundary) but not in the
octahedron (with its boundary) are lazy but entangled. Among these, only the
states at the vertices of tetrahedron are (maximally entangled) pure states.

\section{Summary: a hierarchy diagram}

We explored some 2-qubit states, showed that many states are lazy but
discordant, many states are lazy but entangled, and many states are
disentangled but not lazy. With these investigations, we can surely give a
hierarchy diagram (Figure 1) of 2-qubit states, including lazy states,
disentangled states and zero-discord states.
\begin{figure}[H]
\includegraphics[width=6cm,trim=-50 230 50 0]{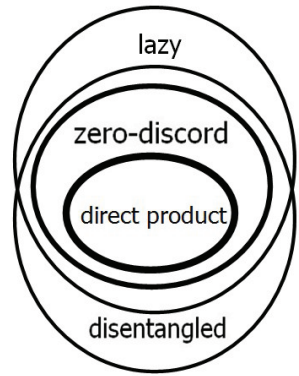}
\caption{laziness-discord-entanglement diagram}
\label{fig1}
\end{figure}

This hierarchy diagram enriches the entanglement-discord hierarchy, then
provides more understandings about the structures of quantum correlations.

This work was supported by the National Natural Science Foundation of
China (Grant No.11347213) and the Research Start-up Foundation for Talents of
Northwest A\&F University of China (Grant No.2013BSJJ041). The author thanks
Zi-Qing Wang and Chang-Yong Liu for helpful discussions.

\end{document}